\documentclass[aps,hidelinks,twocolumn, prl, superscriptaddress,float fix]{revtex4-2}
\usepackage{graphicx}% Include figure files
\usepackage{dcolumn}% Align table columns on decimal point
\usepackage{orcidlink}
\usepackage{caption}
\usepackage{float}
\usepackage{subcaption}
\usepackage{csquotes}
\usepackage{float}
\usepackage{amsfonts}
\usepackage{amsmath}
\usepackage{amsmath,amssymb}
\usepackage[italicdiff]{physics}
\usepackage{bm}
\usepackage{mathtools}
\usepackage{enumitem}
\usepackage{tensor}
\usepackage{wrapfig}
\usepackage{footnote}
\usepackage{here, comment}
\usepackage{bm}% bold math
\begin{document}

%\preprint{YITP-24-94}

\title{Krylov complexity of purification}

\author{Rathindra Nath Das\,\orcidlink{0000-0002-4766-7705}}
\email{das.rathindranath@uni-wuerzburg.de}
\affiliation{Institute for Theoretical Physics and Astrophysics and W\"urzburg-Dresden Cluster of Excellence ct.qmat, Julius-Maximilians-Universit\"at W\"urzburg, 
 Am Hubland, 97074 W\"{u}rzburg, Germany}

\author{Takato Mori\,\orcidlink{0000-0001-6442-875X}}
\email{takato.mori@yukawa.kyoto-u.ac.jp}
\affiliation{Perimeter Institute for Theoretical Physics, Waterloo, Ontario N2L 2Y5, Canada}
\affiliation{Center for Gravitational Physics and Quantum Information, Yukawa Institute for Theoretical Physics, Kyoto University, Kitashirakawa Oiwakecho, Sakyo-ku, Kyoto 606-8502, Japan}

\date{\today}

\begin{abstract}

In quantum systems, purification can map mixed states into pure states and a non-unitary evolution into a unitary one by enlarging the Hilbert space. We establish a connection between the complexities of mixed quantum states and their purification, proposing new inequalities among these complexities. By examining single qubits, two-qubit Werner states, eight-dimensional Gaussian random unitary ensembles, and infinite-dimensional systems, we demonstrate how these relationships manifest across a broad class of systems. We find that the spread complexity of purification of a vacuum state evolving into a thermal state equals the average number of Rindler particles. This complexity is also shown to adhere to the Lloyd{-like} bound, indicating a further relation to the quantum speed limit.
Finally, using \textit{mutual Krylov complexity}, we observe subadditivity of the Krylov complexities, which contrasts with known results from holographic volume complexity. We put forward Krylov mutual complexity as a diagnosis of a potential gravity dual of Krylov complexities.

\end{abstract}

\maketitle

\paragraph*{Introduction.---} 
{Quantum complexity, measuring how complex a quantum state becomes over time, has garnered much attention in various fields, including quantum information, many-body systems, quantum gravity, and cosmology~\cite{Nielsen_2006, nielsen2005geometric, dowling2006geometry, Brown:2018JT, Brown:2017secondlaw, Brown:2019python, Brown:2015action, Parker:2018a, Balasubramanian:2022tpr,Bhattacharyya:2020rpy,Li:2024kfm}. While entanglement effectively characterizes correlations, it fails to capture the full structure of quantum dynamics, motivating complexity as a complementary measure that probes the preparation and evolution of states and operators~\cite{Susskind:2014moa,Stanford:2014complexity,Susskind:2014switchback,Brown:2017secondlaw}. In many-body systems, complexity provides insights into thermalization, chaos, and environmental effects, while in holography it has been tied to black holes, where interior growth corresponds to increasing complexity. This correspondence has led to various proposals for gravity duals of complexity~\cite{Susskind:2014moa,Stanford:2014complexity,Susskind:2014switchback,Brown:2017secondlaw,Brown:2015lvg,Carmi:2017jqz,Susskind:2014rva,Susskind:2018pmk,Yang:2019alh,Engelhardt:2021mju,Couch:2016exn}, making complexity a crucial tool for understanding a black hole interior and information paradox~\cite{Susskind:2014moa,Stanford:2014complexity,Susskind:2014switchback,Brown:2017secondlaw}.}

Among the various definitions of complexity, Krylov operator and state complexity have emerged as powerful tools for studying the dynamical features of both 
unitary and non-unitary 
evolutions~\cite{Parker:2018yvk, Balasubramanian:2022tpr, Caputa:2024vrn, Hornedal:2022pkc, Nandy:2024htc, Bhattacharya:2022gbz, Bhattacharya:2023yec, Bhattacharjee:2022lzy, Bhattacharya:2023zqt, Bhattacharjee:2023uwx, Bhattacharya:2024uxx, Bhattacharya:2024hto, Nandy:2024wwv}. 
It is essential to introduce density matrices for a complete description of the complexity of a quantum state as they are capable of describing mixed states. They are crucial in cases where part of the correlations is classical in nature, when only partial information about the system is available, or when the system is subject to noise and decoherence. A key open question is to coherently understand both the pure and mixed-state complexities 
within the Krylov framework for both unitary and non-unitary evolutions.

In this Letter, we introduce a framework to study Krylov operator complexity of mixed states by relating it to state and operator complexities of their purifications. Unlike earlier approaches embedding the density matrix in a doubled Hilbert space~\cite{Alishahiha_2023}, we employ purification~\cite{Wilde:2011npi,Wilde_2017,Nielsen2012}, mapping a mixed state to a pure state in an enlarged Hilbert space with unitary evolution. Purification has been widely used in entanglement of purification~\cite{Terhal:2002riz} and circuit complexities of purification~\cite{Ag_n_2019,Camargo_2019,Ghodrati:2019hnn,Camargo:2020yfv,Haque:2021kdm,Bhattacharya:2022wlp,Bhattacharyya:2024duw,Bhattacharyya:2021fii,Caceres:2018blh,Caceres:2019pgf,Ruan:2020vze,DiGiulio:2020hlz,DiGiulio:2021fhf,Chapman:2021jbh,Bhattacharya_2021,Yang:2019gce,Wang:2023nes}; our approach extends this by exploiting isometries in purification to connect mixed-state Krylov complexity directly to operator and state complexities, thereby capturing dynamical features that are otherwise missed. We show that for suitable isometries, complexities of purification (CoPs) reproduce the same growth as the original mixed-state complexity.

\begin{figure}
    \centering
    \includegraphics[width=0.5\textwidth]{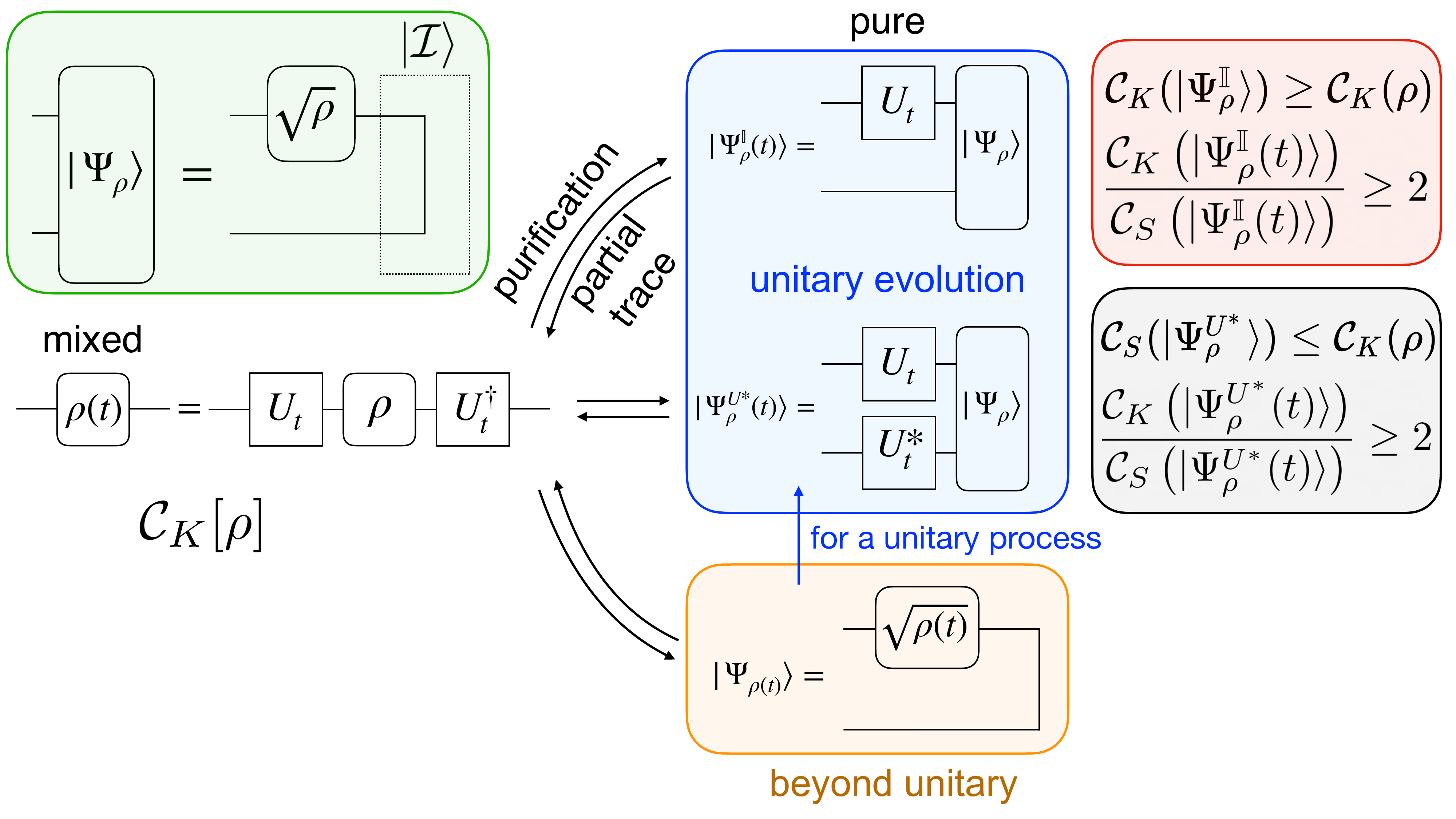}
    \caption{{The top left box shows a canonical purification $\ket{\psi_\rho}$ for a mixed state $\rho$. The central blue and orange boxes show three diferent purification schemes (time-independent, time-dependent, and instantaneous). The rightmost boxes highlight the conjectured bounds for mixed and purification complexities.}
    }
    \label{fig:summary}
\end{figure}

{We examine three purification schemes—time-independent, time-dependent, and instantaneous—each giving distinct complexity growths on the ancilla while describing the same mixed-state evolution. A central result is a set of inequalities between mixed-state complexities and CoPs, as shown in Fig.~\ref{fig:summary}. We demonstrate them for arbitrary single-qubit mixed states, two-qubit Werner states, and the eight-dimensional Gaussian random unitary ensembles. We further verify the conjecture in a non-unitary setting: an infinite-dimensional Gibbs state with increasing temperature, consistent with a quasiparticle picture of the Minkowski vacuum in the Rindler frame. For thermal states, the spread CoP is bounded above by thermodynamic quantities, in line with the Lloyd bound~\cite{Lloyd_2000}. Finally, we introduce \emph{mutual Krylov complexity}, showing it to be subadditive for the thermofield double state—contrasting with holographic complexity in the CV and CV2.0 proposals~\cite{Brown:2015lvg,Carmi:2017jqz,Susskind:2014rva,Susskind:2018pmk,Yang:2019alh,Engelhardt:2021mju,Couch:2016exn}.}

\paragraph*{Definitions.---} 
To evaluate the CoPs, we first discuss three distinct purification schemes. 
Before purification, the operator complexity is the sole option, aside from pure states.
However, once purified, both the state and operator complexities can be examined for the purified states.
The definitions of the state and operator complexities, denoted by $C_{S,K}$, are provided in {the End Matter}.
Given a density matrix $\rho$ of a system $S$, its purification is generically written as
$\ket{\Psi_{\rho}^V}_{SA} = (\rho^{1/2}\otimes V_{R\rightarrow A}) \ket{\mathcal{I}}_{SR}$,
where $V$ is an isometry from $R$ to $A$, i.e. $V^\dagger V = \mathbf{1}_R$, and $\ket{\mathcal{I}}_{SR} = \sum_{n=1}^{\rank\rho} \ket{n}_S \otimes \ket{n}_R$ is an unnormalized EPR state \cite{Wilde:2011npi,Wilde_2017, Nielsen2012}. For 
$\rho$ evolving unitarily by
$U_t=e^{-iHt}$,
the purification evolves as
$\ket{\Psi_\rho^V (t)}= (U_t\rho^{1/2}\otimes V(t))\ket{\mathcal{I}} = (U_t\otimes V(t))\ket{\Psi_\rho}$, where $\ket{\Psi_\rho}:=|\Psi_\rho^{\mathbb{I}}\rangle$.
Note that $\ket{\Psi_\rho}$ is the initial purification with the minimal dimension. It differs from the canonical purification~\cite{Dutta:2019gen} unless $\rho$ is full-rank.
We emphasize that any choice of $V(t)$ results in the same time evolution $\rho\mapsto \rho(t)=U_t\rho U_t^\dagger$ for the initial system.

One possible choice for the isometry is $V(t)=\mathbb{I}$ and we refer to this as \emph{time-independent purification}. 
The time-evolved purified state is denoted by 
\begin{equation}
    \ket{\Psi^\mathbb{I}_\rho (t)}= (U_t\otimes\mathbb{I})\ket{\Psi_\rho}.
    \label{eq:time-indep-pur}
\end{equation}
Alternatively, we can choose a time-dependent isometry $V(t)$. Choosing $V(t)=U_t^\ast$, we define the \emph{time-dependent purification} after time $t$ by
\begin{align}
    \ket{\Psi^{U^\ast}_\rho (t)} = (U_t\rho^{1/2}U_t^\dagger \otimes \mathbb{I})\ket{\mathcal{I}}= (U_t \otimes U_t^\ast) \ket{\Psi_\rho}.
    \label{eq:time-dep-pur}
\end{align}
This is motivated by considering a static state $\rho\propto\mathbb{I}$ and requiring the purification to be also static.
Lastly, \emph{instantaneous purification} is defined as the purification of the density matrix at each moment. 
The purified state at time $t$ is given by
\begin{equation}
    \ket{\Psi_{\rho(t)}}:=(\rho(t)^{1/2}\otimes\mathbb{I})\ket{\mathcal{I}}.
    \label{eq:inst-pur}
\end{equation}
This purification also applies when $\rho$ evolves non-unitarily, in which case a pure state may evolve into a mixed state. Instantaneous purification reduces to time-dependent purification for a unitary evolution. 
 While there are other choices of purifications, the above three schemes are optimal for Krylov CoPs under certain assumptions, unlike circuit ones (see {the End Matter} for more explanations).

The three different purification schemes enable us to define both the Krylov complexity~\eqref{Kcom} of the purified state through its density matrix form and the spread complexity~\eqref{Scom} through its state vector form~\footnote{Notice the difference in normalization with~\cite{Alishahiha_2023}. In~\cite{Alishahiha_2023}, a density matrix is mapped to a vector by the Choi-Jamio\l{}kowski isomorphism, so it needs additional normalization by $\Tr\rho^2$. In our definition of CoPs, we do not need any such normalization once the initial state is normalized as $\Tr\rho=1$.}.

\paragraph*{Inequalities among complexities.---} 
Based on the three types of purifications defined above, we conjecture inequalities for the complexities of mixed states and their purifications, with analytical and numerical evidence provided in the next section~\footnote{For pure states, it is manifest from our definition of CoPs that they reduce to original notions of complexities.}.

\begin{enumerate}
     \item Krylov operator complexity of the mixed state
     is upper bounded by the operator complexity of the time-independent purification and is lower bounded by the state complexity of the time-dependent purification,
     \begin{equation}
         \mathcal{C}_S\qty(\ket{\Psi_{\rho}^{U^\ast}(t)}) \le \mathcal{C}_K(\rho(t))\le \mathcal{C}_K\qty(\ket{\Psi_{\rho}^{\mathbb{I}}(t)})  
         \label{eq:conjecture1}
     \end{equation} 
        where the equality holds in the pure state limit.
        \item The ratio of the state complexity of the time-dependent purification and the operator complexity of the mixed state, $\mathcal{C}_{S}(\ket{\Psi_\rho^{U^\ast}(t)})/\mathcal{C}_K(\rho(t))$, is approximately constant compared to its mean value over time. Its value depends on the purity of $\rho$.

        \item For the same type of purification scheme, the CoPs satisfy $\mathcal{C}_K \qty(\ket{\Psi_\rho^{U^\ast, \mathbb{I}}(t)}) \ge 2 \mathcal{C}_S\qty(\ket{\Psi_\rho^{U^\ast, \mathbb{I}}(t)})$.
        
    \end{enumerate}

\paragraph*{Examples.---}
\begin{figure*}[hbtp]
    \centering
     \begin{subfigure}[b]{0.30\textwidth}
     \centering
         \includegraphics[width=\textwidth]{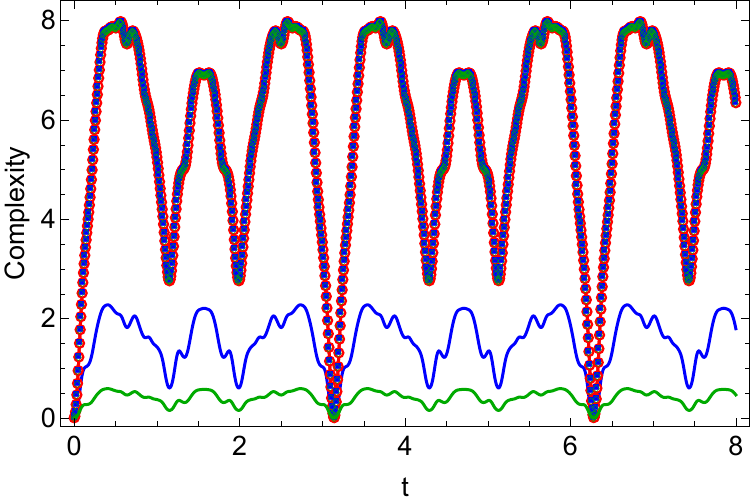}
         \caption{}
         \label{fig:diff_com}
     \end{subfigure}
     \hfill
     \begin{subfigure}[b]{0.32\textwidth}
         \centering
         \includegraphics[width=\textwidth]{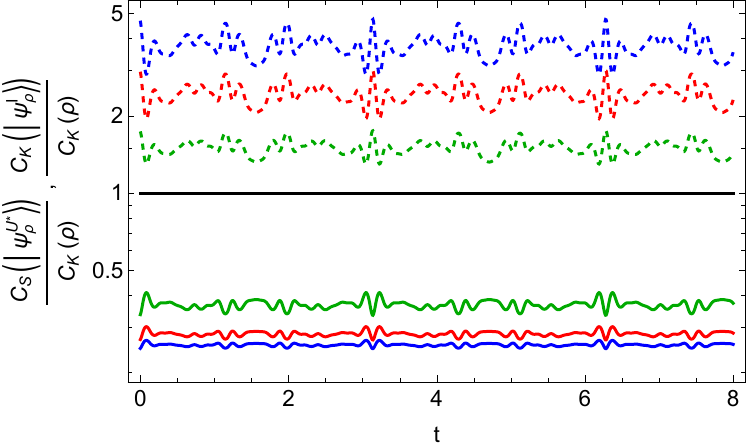}
         \caption{}
         \label{fig:skkk}
     \end{subfigure}
     \hfill
     \begin{subfigure}[b]{0.32\textwidth}
         \centering
         \includegraphics[width=\textwidth]{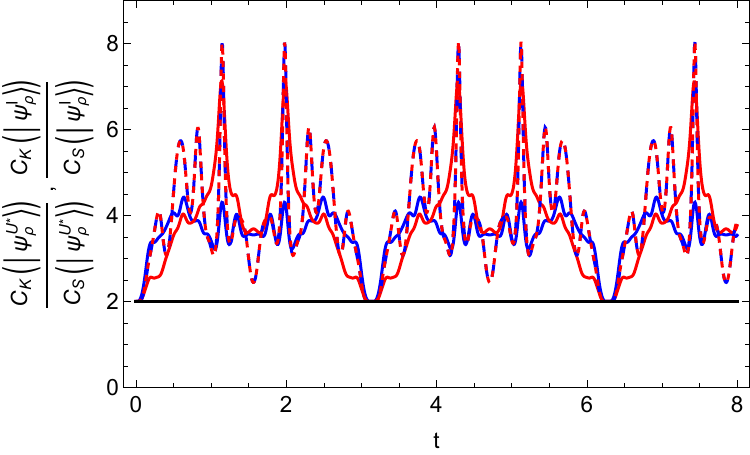}
         \caption{}
         \label{fig:ksks}
     \end{subfigure}
     \\
     \begin{subfigure}[b]{0.30\textwidth}
     \centering
         \includegraphics[width=\textwidth]{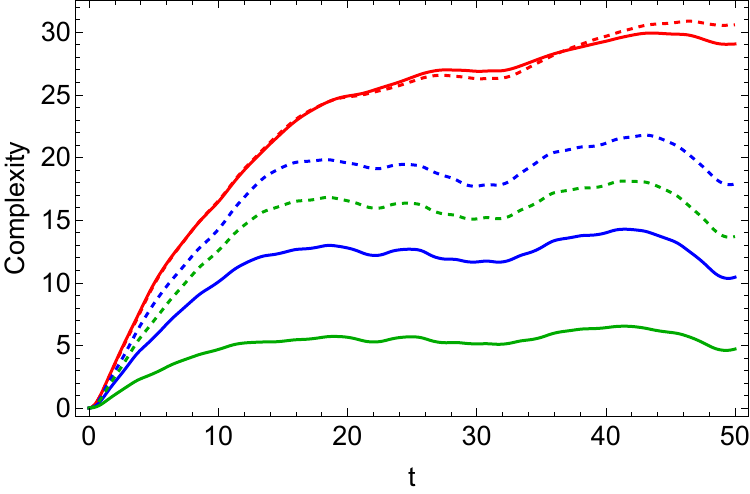}
         \caption{}
         \label{fig:diff_com_rmt}
     \end{subfigure}
     \hfill
     \begin{subfigure}[b]{0.32\textwidth}
         \centering
         \includegraphics[width=\textwidth]{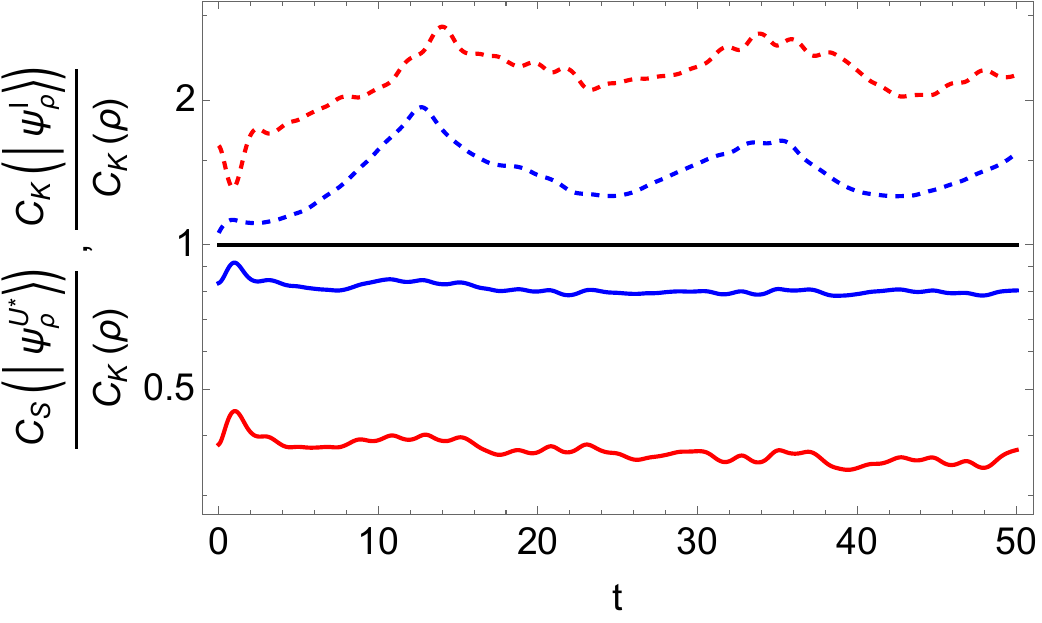}
         \caption{}
         \label{fig:skkk_rmt}
     \end{subfigure}
     \hfill
     \begin{subfigure}[b]{0.32\textwidth}
         \centering
         \includegraphics[width=\textwidth]{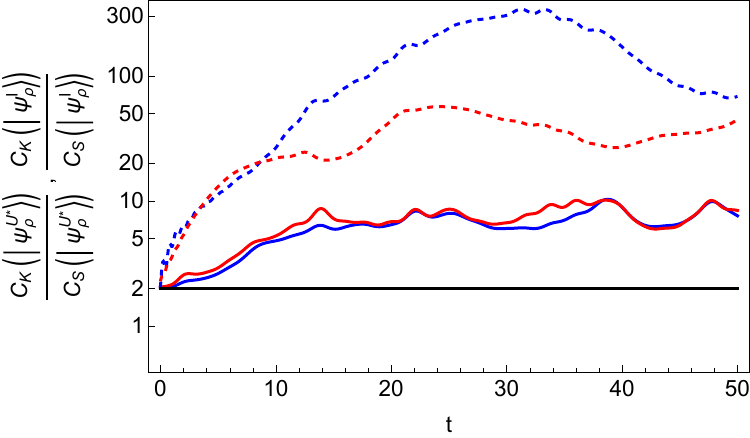}
         \caption{}
         \label{fig:ksks_rmt}
     \end{subfigure}
        \caption{
        {Time dependence of the complexities of two states: [Row 1] Werner states~\eqref{w_state} evolved by the Hamiltonian~\eqref{eq:Hamiltonian_werner} with $r=4,q=15$; [Row 2] Eight-dimensional Gaussian random ensembles with the initial inverse temperature $\beta=1$ (solid line) and $\beta=3$ (dashed line). [Col.~1] CoPs~$\mathcal{C}_K(|\Psi_{\rho}^{\mathbb{I}}(t)\rangle)$ (red), $\mathcal{C}_K(\rho(t))$ (blue), and $\mathcal{C}_S(|\Psi_{\rho}^{U^\ast}(t)\rangle)$ (green). In (a), $p=1/4$ corresponds to a solid curve and $p\rightarrow 1$ corresponds to dotted points. [Col.~2]~Ratio between the operator complexity and CoPs: 
        $\mathcal{C}_K(|\Psi_\rho^{\mathbb{I}}(t)\rangle)/\mathcal{C}_K(\rho(t))$ (dashed curve) and $\mathcal{C}_S(|\Psi_\rho^{U^\ast}(t)\rangle)/\mathcal{C}_K(\rho(t))$ (solid curve). In (b), the ratios with $p=1/4$ (blue), $p=1/3$ (red), $p=1/2$ (green), and $p\rightarrow 1$ (black) are plotted in log scale. [Col.~3]~Ratio $\mathcal{C}_K/\mathcal{C}_S$ for time-independent (dashed) and time-dependent (solid) purifications. In (c), $p=1/4$ (blue), $p\to 1$ (red), with reference line at $2$ (black).}
        %
        % Time dependence of the complexities of the Werner states~\eqref{w_state} with $r=4$ and $q=15$ in the Hamiltonian~\eqref{eq:Hamiltonian_werner}.
        % (a)~$\mathcal{C}_K(|\Psi_{\rho}^{\mathbb{I}}(t)\rangle)$ (red), $\mathcal{C}_K(\rho(t))$ (blue), and $\mathcal{C}_S(|\Psi_{\rho}^{U^\ast}(t)\rangle)$ (green) for different
        % parameters $p=1/4$ (solid curve) and $p\rightarrow 1$ (dotted points). Every complexity  
        % coincides for $p\rightarrow 1$, 
        % and $\mathcal{C}_K(|\Psi_{\rho}^{\mathbb{I}}(t)\rangle)$ overlaps among different $p$. 
        % (b)~Ratio between the operator complexity and CoPs: 
        % $\mathcal{C}_K(|\Psi_\rho^{\mathbb{I}}(t)\rangle)/\mathcal{C}_K(\rho(t))$ (dashed curve) and $\mathcal{C}_S(|\Psi_\rho^{U^\ast}(t)\rangle)/\mathcal{C}_K(\rho(t))$ (solid curve) for  
        % $p=1/4$ (blue), $p=1/3$ (red), $p=1/2$ (green), and $p\rightarrow 1$ (black) in log scale. 
        % (c)~Ratio $\mathcal{C}_K/\mathcal{C}_S$ between the state and operator CoPs for the time-independent purification
        % (dashed curve) and the time-dependent purification
        % (solid curve) for 
        % $p=1/4$ (blue) and $p\rightarrow 1$ (red) with a reference line at $2$ (black). 
        }
        \label{fig:complexity_werner}
\end{figure*}
We first study the CoPs for arbitrary one-qubit mixed states and the two-qubit Werner states and demonstrate the proposed bounds. A one-qubit mixed state under a generic unitary evolution is simple enough that the analytic expressions of its Krylov CoPs can be written down. Leaving the technical details to {the End Matter} and Section II of Supplemental Material, the conjectures in the last section are analytically confirmed.
Let us next study the two-qubit Werner states, {i.e.} %defined as
\begin{equation}
    \rho= \frac{p}{2}(\ket{01} - \ket{10})(\bra{01} - \bra{10}) + \frac{1-p}{4}\mathbb{I}_{4 \times 4}. \label{w_state}
\end{equation}
The parameter $p$ interpolates between the maximally mixed state $\rho\propto\mathbb{I}$ at $p=0$ and a pure, maximally entangled state $\ket{Y}\equiv(\ket{01}-\ket{10})/\sqrt{2}$ at $p=1$.
Additionally, the state is separable for $p \leq \frac{1}{3}$ and entangled otherwise. 
Since the Werner state commutes with the Pauli matrices $\sigma_i\otimes\sigma_i$ for any $i=x,y,z$, let us consider the following Hamiltonian for a nontrivial time evolution:
\begin{equation}
    H = \mathbf{\sigma}_x \otimes \mathbf{\sigma}_y + r \, \mathbf{\sigma}_y \otimes \mathbf{\sigma}_z + q \, \mathbf{\sigma}_z \otimes \mathbf{\sigma}_x,
    \label{eq:Hamiltonian_werner}
\end{equation}
where $r$ and $q$ are free parameters. 
For the time-independent and time-dependent purification, the evolution is governed by $H \otimes \mathbb{I}$ and $H \otimes \mathbb{I} - \mathbb{I} \otimes H^*$, respectively. %While the coefficients of each term in the Hamiltonian~\eqref{eq:Hamiltonian_werner} are {arbitrary,} %not unique, 
%we keep them different from each other to prevent additional symmetries. For Fig.~\ref{fig:complexity_werner}, we choose $r=4$ and $q=15$. 
{The first row of Fig.~\ref{fig:complexity_werner} shows the case with $r=4,q=15$ when $p=1/4$ and $p\rightarrow 1$. We chose $r\neq q$ to prevent {extra} symmetries and we always work with $p<1$ as our primary interest is mixed states.} 
For other choices of parameters, see Section III of the Supplemental Material.

%As we are interested in the mixed-state complexity, we focus on $p<1$. 
As $\rho$ is always full-rank,
the initial state for the CoPs is given by the canonical purification, namely, $|\Psi_\rho\rangle = \sqrt{\frac{1+p}{4}}\ket{YY} + \sqrt{\frac{1-p}{4}}(\ket{II}+\ket{XX}+\ket{ZZ}),$ where $\ket{I},\ket{X},\ket{Y},\ket{Z}$ are the Bell basis~\cite{Nielsen2012}.

{Fig.\ref{fig:diff_com} shows that the operator complexity of the mixed state, $\mathcal{C}_K(\rho(t))$ (blue), is bounded between the CoPs: the time-independent purification $\mathcal{C}_K(\ket{\psi_\rho^\mathbb{I}(t)})$ (red, upper bound) and the time-dependent purification $\mathcal{C}_S(\ket{\psi_\rho^{U^*}(t)})$ (green, lower bound). Solid ($p=1/4$) and dotted ($p\to1$) lines denote different initial states, with recurrence times~\cite{Balasubramanian:2024ghv,Hashimoto:2023swv} being the same. The inequalities~\eqref{eq:conjecture1} are saturated in the pure-state limit $p\to1$, reflecting the doubled Hilbert space via purification\footnote{The equality $\mathcal{C}_K(\rho(t))=\mathcal{C}K(\ket{\Psi\rho^\mathbb{I}(t)})$ for pure states follows from matching autocorrelation functions.}.}

{Fig.~\ref{fig:skkk} shows the time-dependence of the ratio of CoPs to the original mixed-state complexity $\mathcal{C}_K(\rho(t))$. It demonstrates the second conjecture that $\mathcal{C}_K(\rho(t))$ is approximately proportional to the state complexity of the time-dependent purification, with the proportionality constant set by purity. Detailed analysis of fluctuations is provided in Section IV of the Supplemental Material. This supports our initial motivation for CoPs as probes of mixed-state operator complexity, and the comparison to unity further illustrates the inequalities~\eqref{eq:conjecture1}.}

In Fig.~\ref{fig:ksks}, we show that for the same type of purification of the Werner state, $\mathcal{C}_K(t)\ge 2\mathcal{C}_S(t)$ always holds regardless of the choice of initial states.
This property extends a previous observation of $\mathcal{C}_K=2\mathcal{C}_S$ for maximally entangled states~\cite{Caputa:2024vrn} to the inequality for pure states~\footnote{Note that even when the initial state is maximally entangled ($p\rightarrow1$), it is only within the two-dimensional subspace. Thus, this does not mean a contradiction to the observation in~\cite{Caputa:2024vrn}.}.
The conjecture, $\mathcal{C}_K(t)\ge 2\mathcal{C}_S(t)$ can also be shown for general one-qubit pure states. Refer to {the End Matter} for more details.

{To test our conjecture in a larger system, we take a randomly evolving eight-dimensional thermal state, given by}
\begin{equation}
    {\rho(t) = e^{-iHt}\rho_\beta e^{iHt},\quad \rho_\beta=e^{-\beta H_0}/\Tr(e^{-\beta H_0}).}
\end{equation}
{
We draw both $H$ and $H_0$ independently from the Gaussian random unitary ensemble. The CoPs are plotted in the second row of Fig.~\ref{fig:complexity_werner}. The solid line corresponds to $\beta=1$ and the dashed line corresponds to $\beta=3$. The purity is around $0.25$ and $0.65$, respectively.
}

 {
 Although the dimension is too small to capture large-$d$ statistics, Fig.~\ref{fig:diff_com_rmt} shows that all complexities grow approximately linearly up to the Heisenberg time $\sim 8$. The conjectured hierarchy~\eqref{eq:conjecture1} is supported by the plot and by Fig.~\ref{fig:skkk_rmt}. While the ratio $\mathcal{C}_K(\ket{\Psi_\rho^{\mathbb{I}(t)}})/\mathcal{C}_K(\rho(t))$ exhibits strong temporal variation, $\mathcal{C}_S(\ket{\Psi_\rho^{U^\ast}(t)})/\mathcal{C}_K(\rho(t))$ fluctuates only slightly, showing that the spread CoP tracks the operator complexity more closely. Finally, Fig.~\ref{fig:ksks_rmt} confirms our third conjecture, $\mathcal{C}_K \ge 2\mathcal{C}_S$, at all times.
 }

Next, let us apply our framework to a non-unitarily evolving infinite-dimensional diagonal mixed state
\begin{equation}
    \rho(t) = \frac{1}{\cosh^2 (\alpha t)} \sum_{n=0}^\infty \tanh^{2n} r \dyad{n}.
    \label{eq:tmsv-m}
\end{equation}
The evolution interpolates a pure state at $t=0$ and the maximally mixed state at $t\rightarrow\infty$ so it is non-unitary.
The instantaneous purification of~\eqref{eq:tmsv-m} is given by a two-mode squeezed vacuum (TMSV):
\begin{equation}
    \ket{\Psi_{\rho(t)}}=\frac{1}{\cosh r}\sum_{n=0}^\infty (-i)^n \tanh^n(\alpha t) \ket{n}\ket{n}
    \label{eq:tmsv-p}
\end{equation}
as long as $\rho(t)$ is mixed, equivalently $t>0$.
The purified state~\eqref{eq:tmsv-p} can be viewed as $\ket{0}\ket{0}$ driven by a two-mode squeezing Hamiltonian $H = \alpha (a b + a^\dag b^\dag)$ acting on the \emph{enlarged} system, where $a$ ($a^\dag$) and $b$ ($b^\dag$) are the annihilation (creation) operators acting on the system and the ancilla, respectively~\cite{Caves:1985zz, Ou:1992hdj,Haque:2021kdm}. We stress that the time evolution for the state $\rho(r)$ with $r=\alpha t$ is not driven by $H$, but it is a non-unitary, completely positive, trace-preserving map given by tracing out the ancillary system.

Since the evolution is non-unitary, we can consider either the operator complexity of the original mixed state~\eqref{eq:tmsv-m} and the operator/state complexity of its instantaneous purification~\eqref{eq:tmsv-p}. For $\mathcal{C}_K(\rho(t))$, the autocorrelation function is given by $\Tr(\rho(t)\rho(0))$ and for $\mathcal{C}_K(|\Psi_{\rho(t)}\rangle)$, it is given by $[\Tr(\rho^{1/2}(t) \rho^{1/2}(0))]^2$. They are both equal to $\sech^2 (\alpha t)$ \footnote{Interestingly, the scaling of the autocorrelation function in our example $\sech^2 (\alpha t)$ matches with that of the low-$T$ Sachdev-Ye-Kitaev model in the decoupling limit ($q\rightarrow 1$). Here $\alpha$ corresponds to $\pi T$.}. Thus,  their Krylov complexities are also identical. The Krylov complexity with this autocorrelation function is given by~\cite{Parker:2018yvk, Caputa:2021sib}
\begin{equation}
    \mathcal{C}_K(\rho(t)) = \mathcal{C}_K \qty(\ket{\Psi_{\rho(t)}}) = 2 \sinh^2 (\alpha t).
    \label{eq:krylov-tmsv}
\end{equation}

The spread complexity of the instantaneous purification~\eqref{eq:tmsv-p} is easily calculated by the Gaussian nature of the TMSV~\cite{Adhikari:2023evu}. {We find that} %It follows from the Krylov basis being the number basis that
\begin{equation}
    \mathcal{C}_S(\ket{\Psi_{\rho(t)}}) = \ev{n} = \sinh^2(\alpha t).
    \label{eq:spread-tmsv}
\end{equation}
 %Both $\mathcal{C}_K(\rho(t))$ and $\mathcal{C}_S(\ket{\Psi_{\rho(t)}})$ exhibit exponential growth with the same exponent, up to a factor of two.

The above results show that CoPs capture key features of the Krylov complexity of the original non-unitary evolution, reflecting our first conjecture $\mathcal{C}_K(\ket{\Psi}) \ge 2\mathcal{C}_S(\ket{\Psi})$ and $\mathcal{C}_S (\ket{\Psi}) \le \mathcal{C}_K(\rho)$ for the instantaneous purification scheme in the non-unitary evolution. Moreover, computing spread CoP is typically easier than computing the original Krylov complexity. For example, by utilizing the translational invariance of the spread complexity~\cite{Aguilar-Gutierrez:2023nyk}, one can easily find that the spread CoP for the initially mixed state~\eqref{eq:tmsv-m} with $t=t_0$ is given by $\sinh^2(\alpha (t - t_0))$. As no analytical form for the operator complexity is known from its autocorrelation function, it demonstrates an advantage of our proposal of CoP.

It is worth noting that the mixed state~\eqref{eq:tmsv-m} and its purification~\eqref{eq:tmsv-p} can be identified as a Gibbs and TFD state of a harmonic oscillator, respectively~\footnote{This is expected because the two-mode squeezing operation is equivalent to the Bogoliubov transformation, which relates the Rindler vacuum to the Minkowski vacuum.}. The time parameter $t$ is related to the inverse temperature $\beta$ via $\tanh^2 (\alpha t)=e^{-\beta(t) \Delta E}$, where $\Delta E$ is the energy spacing of a harmonic oscillator. 
In other words, the mixed state and its purification after time $t$ can be rephrased as
\begin{equation}
    \rho(t)\propto e^{-\beta(t) H_{\mathrm{HO}}},\quad \ket{\Psi_{\rho(t)}}\propto \sum_{n=0}^\infty e^{-\beta(t)E^{\mathrm{HO}}_n/2} \ket{n}\ket{n},
    \label{eq:tmsv-ho}
\end{equation}
where $H_{\mathrm{HO}}=\sum_n E^{\mathrm{HO}}_n \dyad{n}$ and $E^{\mathrm{HO}}_n=(n+1/2)\Delta E$. We emphasize that while the mixed state is the Gibbs state and the purification is the TFD state with respect to $H_{\mathrm{HO}}$, the purification is \emph{not} driven by $H_{\mathrm{HO}}$ but the two-mode squeezing Hamiltonian $H$.
With this identification, we correspond~\eqref{eq:tmsv-p} with the mode expansion of the two-dimensional Minkowski vacuum of a massless free field in terms of the left (L) and right (R) Rindler basis.
This picture leads to a quasiparticle interpretation of Krylov CoPs, as explained in {the End Matter.} 
An increase in CoPs can be understood as a consequence of exchanging Hawking quanta.

Finally, we highlight a relation between the spread CoP and thermodynamic quantities. The time derivative of the spread CoP~\eqref{eq:spread-tmsv} and the {`modular'} energy $E_{{mod}}=\ev{H_{\mathrm{HO}}}$ and the von Neumann entropy $S$ of~\eqref{eq:tmsv-m} at time $t$, i.e. thermal energy and entropy of {a harmonic oscillator at} inverse temperature $\beta(t)=-(\Delta E)^{-1}\log\tanh^2(\alpha t)$, satisfy~\footnote{{We emphasize here again that the `energy' appearing in the Lloyd bound is not the expectation value of the actual Hamiltonian generating the time evolution. The `modular' energy we call here is the expectation value of a time-independent effective Hamiltonian, modeling the mixed state at each instance of time by a thermal state with a time-dependent temperature.}}
\begin{equation}
    \dot{\mathcal{C}}_S:=\dv{\mathcal{C}_S(|\Psi_{\rho(t)}\rangle)}{t} \le \frac{\alpha}{\Delta E}2E_{{mod}}, \quad \dot{\mathcal{C}}_S\le  \frac{\alpha}{\Delta E} TS.
    \label{eq:dotC}
\end{equation}
{
Since $\Delta E$ is the only characteristic scale, let $\alpha\sim\Delta E$. Then, the bounds become $\dot{\mathcal{C}}_S\lesssim E_{{mod}},\dot{\mathcal{C}}_S\lesssim TS$; the first one saturates at late times, paralleling the Lloyd bound~\cite{Lloyd_2000} for holographic volume complexity~\cite{Brown:2015lvg,Carmi:2017jqz,Susskind:2014rva,Susskind:2018pmk,Yang:2019alh,Engelhardt:2021mju,Yang:2019gce}. We note that the bound is also reminiscent of the quantum speed limit~\cite{Hornedal:2022pkc, Bhattacharya:2024uxx, Srivastav:2024apk}.
}

Despite the similarity of the Lloyd bound, we find that the operator complexity and CoPs in the Krylov formalism differ from the holographic ones based on CV/CV2.0 proposals. To demonstrate this, let us introduce the mutual Krylov complexity for a bipartite state $\rho_{AB}$ as
\begin{equation}
\Delta\mathcal{C}(A:B) = \mathcal{C}(\rho_A(t)) + \mathcal{C}(\rho_B(t)) - \mathcal{C}(\rho_{AB}(t)). \label{eq:MKC}
\end{equation}
This quantity itself has been introduced in the context of the Nielsen and holographic complexities~\cite{Alishahiha:2018lfv,Caceres:2019pgf}.
Possible choices for the complexity measure $\mathcal{C}$ are the Krylov operator complexity, the subsystem complexity (see {the End Matter} for its definition), and the operator/state CoPs.
For the TFD case, we have $\Delta \mathcal{C}(L:R)=2\mathcal{C}(\rho(t))-\mathcal{C}(|\Psi_{\rho(t)}\rangle)$. The mutual Krylov complexity $\Delta \mathcal{C}(L:R)$ equals $2 \sinh^2(\alpha t)$ for the operator complexity/CoP ($\mathcal{C}=\mathcal{C}_K$) in~\eqref{eq:MKC}, and $\sinh^2(\alpha t)$ for the state complexity/CoP ($\mathcal{C}=\mathcal{C}_S$) in~\eqref{eq:MKC}~\footnote{Note that the instantaneous purification of the TFD state equals itself.}.
In all cases, the operator complexity and state/operator CoPs show subadditivity, $\Delta\mathcal{C}(L:R)\ge 0$, in contrast to superadditivity of holographic volume complexity~\cite{Caceres:2019pgf, Ag_n_2019, Ruan:2021wep, C_ceres_2019,Caceres:2018blh}. As the present calculation is based on the free field $E_n\propto n$, 
further computations in holographic conformal field theory (CFT) are needed to confirm the discrepancy.

\paragraph*{Summary and outlook.---} 

In summary, we introduced mixed-state complexity within the Krylov formalism via purification and established bounds between mixed-state complexity and CoPs, verified in one- and two-qubit systems and the eight-dimensional Gaussian random ensembles. The spread complexity of time-dependent purification closely tracks the mixed-state operator complexity while being easier to compute, and our framework also extends to infinite-dimensional thermal states, where the inequalities are satisfied and even saturated. We showed that while spread complexity obeys a Lloyd-like bound, Krylov complexities for free field theories exhibit subadditivity, differing from complexity=volume proposals in holography~\cite{Rabinovici:2023yex,Balasubramanian:2024lqk}; this contrasts with strongly coupled systems such as DSSYK, highlighting the role of integrability, a question we leave for future work. Although a systematic analysis of genuinely large many-body systems is beyond the scope of this work, as a first step in this direction we are currently investigating $\mathcal{C}_K(\rho(t))$, $\mathcal{C}_S(|\Psi\rho^{U^\ast}(t)\rangle)$, and $\mathcal{C}_K(|\Psi\rho^{\mathbb{I}}(t)\rangle)$ in the large-$N$ limit of random matrix theory, where we find analytic agreement of the first 20 Lanczos coefficients among them, providing preliminary evidence that our conjecture can be saturated in a solvable large-$N$ setting~\cite{Mori:2025RMTKrylov}. These results suggest that our bounds may serve as practical diagnostics of wave-function spreading and information transport in many-body systems (e.g. Lieb–Robinson bounds) and help clarify the link between Krylov complexity, Lloyd-type bounds, and quantum speed limits. In particular, it is interesting to ask whether a Lloyd-like bound can be found generically in nonequilibrium thermodynamics in the light of the modular Hamiltonian.

%{In summary, we introduced mixed-state complexity within the Krylov formalism via purification and established bounds between mixed-state complexity and CoPs, verified in one- and two-qubit systems and the 8-dimensional Gaussian random ensemble. The spread complexity of time-dependent purification closely tracks the mixed-state operator complexity while being easier to compute, and our framework also extends to infinite-dimensional thermal states, where the inequalities are satisfied and even saturated. We showed that while spread complexity obeys a Lloyd-like bound, Krylov complexities for free field theories exhibit subadditivity, differing from complexity=volume proposals in holography~\cite{Rabinovici:2023yex,Balasubramanian:2024lqk}; this contrasts with strongly coupled systems such as DSSYK, highlighting the role of integrability, a question we leave for future work. More broadly, our bounds could be tested in many-body systems, potentially constraining wave-function spreading (e.g. via Lieb–Robinson bounds) and deepening the link between Krylov complexity, the Lloyd bound, and quantum speed limits. In particular, it is interesting to ask the Lloyd-like bound can be found generically in nonequilibrium thermodynamics in the light of modular Hamiltonian.}

 \begin{acknowledgments}
 \vspace{2em}
 \noindent The authors would like to thank Souvik Banerjee, Pablo Basteiro,  Aranya Bhattacharya, Giuseppe Di Giulio, Johanna Erdmenger, Rob Myers, and Shang-Ming Ruan for useful discussions and comments. R.N.D.~is supported by Germany's Excellence Strategy through the W\"urzburg-Dresden Cluster of Excellence on Complexity and Topology in Quantum Matter - ct.qmat (EXC 2147, project-id 390858490), and by the Deutsche Forschungsgemeinschaft (DFG)  through the Collaborative Research centre \enquote{ToCoTronics}, Project-ID 258499086-SFB 1170. This research was also supported in part by the Perimeter Institute for Theoretical Physics. Research at Perimeter Institute is supported by the Government of Canada through the Department of Innovation, Science and Economic Development and by the Province of Ontario through the Ministry of Research, Innovation and Science. This work was supported by JSPS KAKENHI Grant Number 23KJ1154, 24K17047.
 \end{acknowledgments}

 ~

\section*{End Matter}

\paragraph*{Definition of Krylov and spread complexity---}
We start by reviewing the definitions of the Krylov complexity for a density matrix and the spread complexity for a pure state. For a detailed discussion, refer to Section I of the Supplemental Material. Let us consider a density matrix $\rho(t)$ that follows the Liouville-von Neumann equation, $\rho(t) = e^{-i H t} \,\rho(0)\, e^{i H t}$~\footnote{Note that the sign of the evolution arises from the state evolution, contrasting with the Heisenberg operator evolution commonly considered in the Krylov operator complexity.}. Using the Baker-Campbell-Hausdorff formula $\rho(t)$ can be written as the sum of nested commutators of $H$ and $\rho(0)$. By making a choice of the inner product, say, $\left(\rho_1 | \rho_2\right) := \text{Tr}\left[\rho_1^{\dagger} \rho_2 \right]$ and using the Lanczos algorithm on the set of operators that are the nested commutators $\comm{H}{\comm{H,\cdots [H}{\rho(0)]\cdots}}$, we obtain the Krylov basis $|\phi_n)$. The time-dependent density matrix $\rho(t)$ can be expanded in the Krylov basis $|\rho(t)) = \sum_n i^n \varphi_n(t) |\phi_n)$, where the probability amplitudes $\varphi_n(t)$ satisfy $\sum |\varphi_n(t)|^2 =1$. This leads to the Krylov operator complexity $\mathcal{C}_{K}$ of the operator $\rho(t)$,
\begin{align}\label{Kcom}
\begin{split}
\mathcal{C}_{K}(t) := \sum_n n\left|\varphi_n(t)\right|^2.
\end{split}
\end{align}
Krylov complexity can alternatively be calculated from the autocorrelation function $G_K(t)=(\rho(t)|\rho(0))$~\cite{Nandy:2024htc}.

The spread complexity (also known as the Krylov state complexity) is the optimal measure of complexity that quantifies the spread of a pure state as it evolves~\cite{Balasubramanian:2022tpr}. Starting with an initial state $|\psi(0)\rangle$ along with the Schr\"odinger evolution, $|\psi(t)\rangle= e^{-i H t} |\psi(0)\rangle,$ the Krylov basis $|K_n\rangle$ can be built from the set of states $\{H^n|\psi(0)\rangle, n\in \mathbb{Z}\}$ by using the Lanczos algorithm as reviewed in Section I in the Supplemental Material. Expanding the time-evolved state in the Krylov basis leads to $|\psi(t)\rangle = \sum_{n} \psi_n(t) |K_n\rangle$. 
Here, $|\psi_n(t)|^2$ represents the probability of the state being in the $n$-th Krylov basis element $|K_n\rangle$ at time $t$, with total probability $\sum_{n} |\psi_n(t)|^2 = 1$. The spread complexity $\mathcal{C}_S$ of a pure state $\ket{\psi(t)}$ is defined as the average position in the Krylov space of states
\begin{equation} \label{Scom}
\mathcal{C}_S(t) := \sum_{n} n|\psi_n(t)|^2.
\end{equation}

~

 \paragraph*{CoPs and the choice of isometry---}
We do not require minimization over purification unlike previous proposals of the circuit CoP, however, our definition of CoPs is optimal under certain assumptions. For time-independent isometry, the Krylov/spread complexities are unaffected by the isometry choice, as its dependence cancels out in the autocorrelation function. For time-dependent isometry, we can always fix the dimension of the Hilbert space as time evolves by preparing a sufficiently large number of ancilla. This implies
the isometry can be decomposed as $V(t)=W(t)V_0$, where $V_0$ is time-independent, and $W(t)$ is a time-dependent unitary. For the CoP to approximate the original complexity well, we need CoP to be zero when the original complexity is zero (no time evolution), requiring $W(t) V_0 = W_0 U_t^\ast$, with $W_0$ an arbitrary time-independent isometry. As the upper bound of complexity grows with dimension~\cite{Rabinovici:2020operator}, the smallest complexity is likely to be achieved by the smallest purification. Setting $V_0=W_0=\mathbb{I}$ gives the time-dependent purification~\eqref{eq:time-dep-pur}. 
A similar argument leads to the instantaneous purification~\eqref{eq:inst-pur}.

~

\paragraph*{One-qubit CoPs---} We here discuss the CoPs for arbitrary one-qubit mixed states under a generic unitary evolution and illustrate the proposed bounds. 
An arbitrary one-qubit mixed state can be described as $\rho = V\rho_0 V^\dag$, where $\rho_0=\mathrm{diag}(p_1,p_2)$ with $0 < p_1=1-p_2<1$ and $V$ is an arbitrary one-qubit unitary parametrized as
\begin{equation}
    V=
    \begin{pmatrix}
        \cos\theta/2 & - e^{i\lambda} \sin \theta/2 \\
        e^{-i\phi} \sin\theta/2 & e^{i(\lambda-\phi)}\cos\theta/2
    \end{pmatrix}
    .
\end{equation}
In the energy eigenbasis, any Hamiltonian is given by $H=\mathrm{diag}(E_1,E_2)$. Following the moment method as employed in~\cite{Caputa:2024vrn}, we find $C_K^{\mathbb{I}},C_K,C_S^{U^\ast}$ are expressed by a common function $C[\mu]=\mu \qty(\sin^2\tau + 8(1-\mu) \sin^4\frac{\tau}{2})$ as
\begin{equation}
    C_K^{\mathbb{I}}=C[\mu^{\mathbb{I}}],\quad
    C_K=C[\mu_\ast],\quad
    C_S^{U^\ast}=C[\mu^{U^\ast}]
\end{equation}
with $\mu^{\mathbb{I}}\equiv \frac{1-(\Delta p)^2 \cos^2\theta}{2}$, $\ \mu_\ast\equiv \frac{(\Delta p)^2\sin^2\theta}{1+(\Delta p)^2}$, $\mu^{U^\ast} \equiv \frac{(\Delta\sqrt{p})^2\sin^2\theta}{2}$
in terms of $\Delta p = p_1-p_2$ and $\Delta\sqrt{p}=\sqrt{p_1}-\sqrt{p_2}$. Using these analytical expressions, we confirm our conjecture~\eqref{eq:conjecture1} for arbitrary one-qubit mixed states and evolution. We also confirm that $C_K\ge 2C_S$ for the CoPs  and the ratio $\mathcal{C}_{S}(\ket{\Psi_\rho^{U^\ast}(t)})/\mathcal{C}_K(\rho(t))$ is approximately constant compared to its mean value over time, which depends on the purity of $\rho$. It is worth noting that $C_K^{\mathbb{I}}$ has a vanishing purity dependence when $\theta=\pi/2$, which means the basis of the time evolution is orthogonal to the diagonalizing basis for the initial state. This is unlike $C_K$ or $C_S^{U^\ast}$, whose purity dependence vanishes only when there is no time evolution ($\theta= 0,\pi$). For further details, see Section II 
of the Supplemental Material.  

~

\paragraph*{The inequality between Krylov and spread complexity of pure one-qubit states---}
To further support the conjecture $C_K\ge 2C_S$, let us consider a pure one-qubit state undergoing an arbitrary unitary evolution.
As discussed in~\cite{Caputa:2024vrn}, the Krylov operator complexity of a pure qubit $\rho=\dyad{\psi}$ such that
\begin{equation}
    \ket{\psi} = \cos\theta \ket{E_1}+\sin\theta e^{i\phi} \ket{E_2},
\end{equation}
under an arbitrary time evolution generated by the Hamiltonian $H= E_1\dyad{E_1}+E_2 \dyad{E_2}$ is given by
\begin{equation}
    \mathcal{C}_K(\ket{\psi(t)}) = \frac{1}{2} \sin^2(2\theta) \qty(\sin^2\tau + 2(3+\cos(4\theta))\sin^4\frac{\tau}{2}),
\end{equation}
where we defined a dimensionless time parameter $\tau = \Delta E\, t$.

The spread state complexity of $\rho$ is given by
\begin{equation}
    \mathcal{C}_S(\ket{\psi(t)}) = \sin^2(2\theta) \sin^2\frac{\tau}{2}.
\end{equation}

Its ratio is calculated as
\begin{equation}
    \frac{\mathcal{C}_K(\ket{\psi(t)})}{2\mathcal{C}_S(\ket{\psi(t)})} = 1+ \cos^2(2\theta) \sin^2\frac{\tau}{2} \ge 1.
\end{equation}
This confirms the inequality $\mathcal{C}_K\ge 2\,\mathcal{C}_S$ for an arbitrary one-qubit pure state under an arbitrary unitary evolution.

~

\paragraph*{Quasiparticle interpretation Krylov and spread complexity of TFD evolution---}
Based on the identification between the TMSV and the TFD state~\eqref{eq:tmsv-ho}, let us give a quasiparticle interpretation to~\eqref{eq:spread-tmsv}. The complexity of the vacuum state $\mathcal{C}_S(|\Psi_{\rho(t)}\rangle)$ grows by exchanging the Rindler particles between two Rindler wedges. Then, the growth rate~\eqref{eq:spread-tmsv} is naturally given by the average number of the exchanged quasiparticles. The same argument applies for a perturbation falling into a black hole~\footnote{Note that the mapping between time and the temperature $\tanh^2(\alpha t)=\exp(-\beta(t) \Delta E)$ agrees with that of the perturbation in the late time~\cite{Parker:2018yvk, Caputa:2021sib}.}. By exchanging one Hawking quantum between the exterior and its interior partner, the state thermalizes and the spread complexity increases by one. The quasiparticle picture also explains the factor $2$ of $\mathcal{C}_K(\rho(t))$ in~\eqref{eq:krylov-tmsv}. Since the original mixed state only sees one Rindler wedge, both the absorption and emission happen simultaneously. This increases the complexity by two per unit of time.

~

\paragraph*{Subsystem spread complexity---}
Following \cite{Alishahiha_2023}, we can define another complexity for the subsystem than CoPs. Provided the subsystem spread operator given by $K_L = \Tr_R \sum_n n\dyad{n_L n_R} = \sum_n n\dyad{n}_L$,
the subsystem spread complexity for $L$ is given by
\begin{equation}
    \mathcal{C}^L_S(\rho(t)) = \Tr(K_L \rho_L(t)). \label{eq:sub_TFD}
\end{equation}
For the infinite-dimensional diagonal state~\eqref{eq:tmsv-m}, it equals $\sinh^2 (\alpha t)$.
The equality with the spread CoP is explained in the quasiparticle picture as follows. Because the spread operator is only defined in one subsystem, the effect of the incoming quanta is excluded, and the subsystem complexity grows by the outgoing quanta. 

\setcounter{secnumdepth}{1}

\onecolumngrid
\renewcommand{\theequation}{S.\arabic{equation}}
\setcounter{equation}{0}

\section*{Supplemental Material}
\section{Krylov and Spread complexity}\label{app:Krylov_Spread_com}

\subsection{Spread Complexity \label{sec:spread}}
In this subsection, we summarize the key concepts necessary to measure the complexity associated with the spreading of a quantum state and the evolution of an operator in the Krylov space for systems governed by Hermitian Hamiltonians~\cite{Nandy:2024htc}. For state complexity, we begin with a pure state, with unitary dynamics ensuring that the state remains pure during evolution. For such a quantum system governed by the Hamiltonian $H$, the spread (state) complexity can be defined. Consider a time-evolved state $|\psi(t)\rangle = e^{-iHt}|\psi(0)\rangle$. This can be written as a linear combination of
\begin{equation}
|\psi\rangle,~ H|\psi\rangle,~ H^2|\psi\rangle,~ \cdots,
\label{eq:basis state}
\end{equation}
where $|\psi(0)\rangle$ is denoted as $|\psi\rangle$. The subspace $\mathcal{H}_{|\psi\rangle}$ spanned by \eqref{eq:basis state} is known as the Krylov space. Using the natural inner product, we can orthonormalize \eqref{eq:basis state} using the Lanczos algorithm:
\begin{itemize}
\item[1.] $b_0 \equiv 0$,
\item[2.] $|K_0\rangle \equiv |\psi(0)\rangle, a_0 = \langle K_0|H|K_0\rangle$
\item[3.] For $n \geq 1$: $|\mathcal{A}_n\rangle = (H - a_{n-1})|K_{n-1}\rangle - b_{n-1}|K_{n-2}\rangle$
\item[4.] Set $b_n = \sqrt{\langle \mathcal{A}_n|\mathcal{A}_n\rangle}$
\item[5.] If $b_n = 0$, stop; otherwise, set $|K_n\rangle = \frac{1}{b_n}|\mathcal{A}_n\rangle, a_n = \langle K_n|H|K_n\rangle$, and repeat step 3.
\end{itemize}
If $\mathcal{D}\equiv \dim \mathcal{H}_{|\psi\rangle}$ is finite, the Lanczos algorithm concludes with $b_{\mathcal{D}} = 0$. The resulting orthonormal basis $\{|K_n\rangle\}_{n=0}^{\mathcal{D}-1}$ is called the Krylov basis. Note that there are two sets of Lanczos coefficients $\{a_n\}$ and $\{b_n\}$ in this context.
Expressing $|\psi(t)\rangle$ in terms of the Krylov basis yields
\begin{equation}
|\psi(t)\rangle = \sum_{n=0}^{\mathcal{D}-1} \psi_n(t) |K_n\rangle,
\label{eq:state in the Krylov basis}
\end{equation}
and substituting \eqref{eq:state in the Krylov basis} into the Schr\"odinger equation, we obtain
\begin{equation}
i \dot{\psi}_n(t) = a_n \psi_n(t) + b_{n+1} \psi_{n+1}(t) + b_n \psi_{n-1}(t).
\label{eq:Krylov chain for state}
\end{equation}
The initial condition is $\psi_n(0) = \delta_{n0}$ by definition. Importantly, the weights $\psi_n$ of each Krylov basis element $|K_n\rangle$ can be interpreted as a wave function with support on a semi-infinite chain.

The (Krylov) state complexity is 
the extent to which the state has spread along the chain, or equivalently, the number of Krylov basis elements it encompasses. More precisely, the spread complexity of the state $|\psi(t)\rangle$ is defined as
\begin{equation}
\mathcal{C}_S(t) \equiv \sum_{n=0}^{\mathcal{D}-1} n |\psi_n(t)|^2.
\label{eq:Krylov state complexity}
\end{equation}
As intuitively expected, this notion of complexity serves as a probe for chaos. Note, however, that chaos will manifest differently in state spread complexity compared to (operator) Krylov complexity, as suggested by the conjecture in~\cite{Parker:2018yvk}. Typically, the spread of states in a Hilbert space, as opposed to operators, will display a distinct signature for chaos.

%%%%%%%%%%%%%%%%%%%%%%%%%%%%%%%%%%%%%%%%%%%
\subsection{Krylov Operator Complexity}

In this subsection, we review Krylov operator complexity~\cite{Nandy:2024htc}. Consider a quantum system evolving under the Hamiltonian $H$. In the Heisenberg picture, the Krylov complexity for an operator $\mathcal{O}$ is defined as follows. The time-evolved operator $\mathcal{O}(t) = e^{iHt} \mathcal{O}(0) e^{-iHt}$ can be expanded as
\begin{equation}
\mathcal{O}(t) = \sum_{n=0}^\infty \frac{(it)^n}{n!} \mathcal{L}^n \mathcal{O}(0),
\label{eq:BCH}
\end{equation}
where $\mathcal{L}$ is the Liouvillian superoperator, given by $\mathcal{L} = [H, \cdot]$. This is a superposition of the following operators:
\begin{equation}
\mathcal{O}, \mathcal{L}\mathcal{O}, \mathcal{L}^2\mathcal{O}, \cdots,
\label{eq:nested commutators}
\end{equation}
where $\mathcal{O}$ represents $\mathcal{O}(0)$. The space $\mathcal{H}_{\mathcal{O}}$ spanned by \eqref{eq:nested commutators} is termed the Krylov space associated with $\mathcal{O}$. By introducing an inner product between operators $\mathcal{O}_1$ and $\mathcal{O}_2$ as, for example,
\begin{equation}
(\mathcal{O}_1 | \mathcal{O}_2) \equiv {\rm Tr} \big[ \mathcal{O}_1^\dagger \mathcal{O}_2 \big],
\label{eq:inner product}
\end{equation}
we can construct an orthonormal basis for $\mathcal{H}_{\mathcal{O}}$ using the Lanczos algorithm:
\begin{itemize}
\item[1.] $b_0 \equiv 0, \quad \mathcal{O}_{-1} \equiv 0$
\item[2.] $\mathcal{O}_0 \equiv \mathcal{O} / |\mathcal{O}|$, where $|\mathcal{O}| \equiv \sqrt{(\mathcal{O} | \mathcal{O})}$
\item[3.] For $n \geq 1$: $\mathcal{A}_n = \mathcal{L}\mathcal{O}_{n-1} - b_{n-1}\mathcal{O}_{n-2}$
\item[4.] Set $b_n = |\mathcal{A}_n|$
\item[5.] If $b_n = 0$, stop; otherwise, set $\mathcal{O}_n = \mathcal{A}_n / b_n$ and repeat step 3.
\end{itemize}
In finite-dimensional systems, the Lanczos algorithm terminates with $b_{K_{\mathcal{O}}} = 0$, where $K_{\mathcal{O}} \equiv \dim \mathcal{H}_{\mathcal{O}}$. This yields the orthonormal basis $\{\mathcal{O}_n\}_{n=0}^{K_{\mathcal{O}}-1}$ termed the Krylov basis, and positive numbers $\{b_n\}$ known as the Lanczos coefficients.
Expanding the Heisenberg operator $\mathcal{O}(t)$ in terms of the Krylov basis gives
\begin{equation}
\mathcal{O}(t) = \sum_{n=0}^{K_{\mathcal{O}}-1} i^n \varphi_n(t) \mathcal{O}_n,
\label{eq:operator in the Krylov basis}
\end{equation}
where $\varphi_n(t)$ satisfies the normalization condition
\begin{equation}
\sum_{n=0}^{K_{\mathcal{O}}-1} |\varphi_n(t)|^2 = |\mathcal{O}|^2 = 1
\end{equation}
after correctly normalizing the initial operator $\mathcal{O}$.
Substituting \eqref{eq:operator in the Krylov basis} into the Heisenberg equation leads to
\begin{equation}
\dot{\varphi}_n(t) = b_n \varphi_{n-1}(t) - b_{n+1} \varphi_{n+1}(t),
\label{eq:Krylov chain for operator}
\end{equation}
where the dot represents the derivative with respect to time. The initial condition is $\varphi_n(0) = \delta_{n0} |\mathcal{O}|$ by definition, which simplifies to $\varphi_n(0) = \delta_{n0}$ after normalizing the operator. The Krylov complexity for the operator $\mathcal{O}$ is defined as
\begin{equation}
\mathcal{C}_{\mathcal{O}}(t) \equiv \sum_{n=0}^{K_{\mathcal{O}}-1} n |\varphi_n(t)|^2.
\label{eq:def Krylov operator complexity}
\end{equation}
The operator $\mathcal{O}_n$ includes the nested commutator $\mathcal{L}^n \mathcal{O}$, which generally becomes more complex as $n$ increases. Thus, Krylov operator complexity quantifies the number of nested commutators in the Heisenberg operator $\mathcal{O}(t)$. Both Krylov and the spread complexity can also be computed from the autocorrelation function and the return amplitude, respectively. For the details, refer to the following review~\cite{Nandy:2024htc}.

\section{Complexities of purification of arbitrary one-qubit mixed states under an arbitrary unitary evolution}
In this section, we compute the spread/Krylov complexities of purification for generic one-qubit mixed states and verify our conjectures. A general one-qubit density matrix $\rho$ can be obtained by rotating a diagonal mixed state by a generic one-qubit unitary $V$:
\begin{equation}
    \rho=V\rho_0 V^\dag
    =
    \begin{pmatrix}
        p_1 \cos^2\frac{\theta}{2} + p_2 \sin^2\frac{\theta}{2} & e^{i\phi} \frac{p_1-p_2}{2}\sin\theta\\
        e^{-i\phi} \frac{p_1-p_2}{2} \sin\theta & p_1 \sin^2\frac{\theta}{2} + p_2 \sin^2\frac{\theta}{2}
    \end{pmatrix}
    ,
    \quad
    V=
    \begin{pmatrix}
        \cos\theta/2 & - e^{i\lambda} \sin \theta/2 \\
        e^{-i\phi} \sin\theta/2 & e^{i(\lambda-\phi)}\cos\theta/2
    \end{pmatrix}
    \label{eq:notation-1q}
\end{equation}
where $\rho_0=\mathrm{diag}(p_1,p_2)$ and the global phase in $V$ is already removed as it does not change $\rho$.
From the normalization of $\rho$, we have $p_1+p_2=1$. The rotating angles $\theta,\phi,\lambda$ generate coherence.
Choosing the basis to be the energy eigenstates, the Hamiltonian is a diagonal matrix, namely,
\begin{equation}
    H=
    \begin{pmatrix}
        E_1 & 0\\
        0 & E_2
    \end{pmatrix}
    , \quad E_1\le E_2.
\end{equation}
The time evolution of the state is given by $U_t=e^{-iHt}$.
We denote the energy spacing by $\Delta E=E_2-E_1$ and define the dimensionless time $\tau = t\Delta E$ for convenience.

Following the calculation of the Krylov complexity of mixed states~\cite{Caputa:2024vrn}, we can similarly compute the Krylov and spread complexities of purification of $\rho$. The unnormalized autocorrelation functions are defined as
\begin{align}
    G_S^{\,\mathbb{I}}(t) &= \braket{\Psi^{\mathbb{I}}_{\rho}(t)}{\Psi^{\mathbb{I}}_{\rho}(0)}=\Tr(\rho U^\dag_t)\\
    G_K^{\,\mathbb{I}}(t) &= \Tr \qty[ \dyad{\Psi^{\mathbb{I}}_{\rho}(t)}\dyad{\Psi^{\mathbb{I}}_{\rho}(0)}]=\abs{\Tr(\rho U^\dag_t)}^2\\
    G_K(t) & = \Tr(\rho(t) \rho)\\
    G_S^{U^\ast}(t) &= \braket{\Psi^{U^\ast}_{\rho}(t)}{\Psi^{U^\ast}_{\rho}(0)}=\Tr(\rho^{1/2}(t) \rho^{1/2})\\
    G_K^{U^\ast}(t) &= \Tr \qty[ \dyad{\Psi^{U^\ast}_{\rho}(t)}\dyad{\Psi^{U^\ast}_{\rho}(0)}]=\abs{\Tr(\rho^{1/2}(t) \rho^{1/2})}^2,
\end{align}
where $\rho^{1/2}(t)=U_t V\rho_0^{1/2}V^\dag U_t^\dag$.
Using the notations~\eqref{eq:notation-1q}, they are calculated as
\begin{align}
    G_S^{\,\mathbb{I}}(t) &= \qty(p_1 \cos^2 \frac{\theta}{2}+p_2 \sin^2\frac{\theta}{2})e^{iE_1 t} + \qty(p_1 \sin^2 \frac{\theta}{2}+p_2 \cos^2\frac{\theta}{2})e^{iE_2 t}\\
    G_K^{\,\mathbb{I}}(t) &= 1-\qty(1-(\Delta p)^2 \cos^2\theta) \sin^2\frac{\tau}{2} \\
    G_K(t) & = p_1^2+p_2^2 - (\Delta p)^2 \sin^2\theta \sin^2\frac{\tau}{2}\\
    G_S^{U^\ast}(t) &= 1-(\Delta \sqrt{p})^2\sin^2\theta \sin^2\frac{\tau}{2} \\
    G_K^{U^\ast}(t) &= \qty(1-(\Delta \sqrt{p})^2\sin^2\theta \sin^2\frac{\tau}{2})^2,
\end{align}
where $\Delta p\equiv p_1-p_2$ and $\Delta \sqrt{p}\equiv \sqrt{p_1}-\sqrt{p_2}$. Note that they do not depend on the rotation angles except for $\theta$.

\begin{figure*}[hbtp]
    \centering
     \begin{subfigure}[b]{0.30\textwidth}
     \centering
         \includegraphics[width=\textwidth]{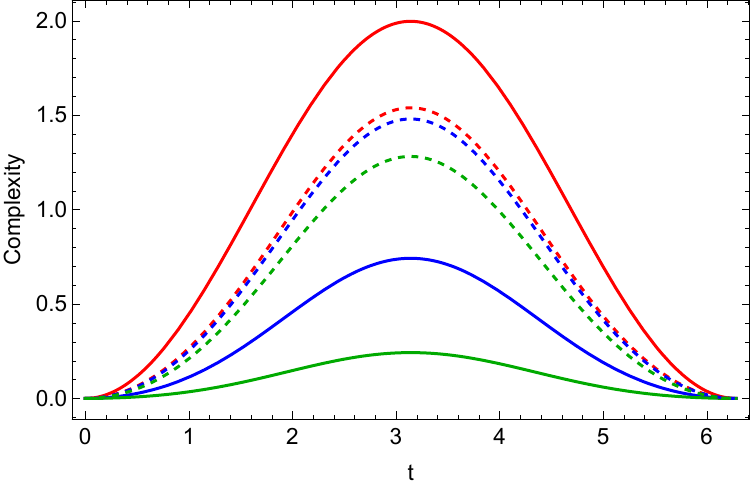}
         \caption{}
         \label{fig:diff_com1q}
     \end{subfigure}
     \hfill
     \begin{subfigure}[b]{0.32\textwidth}
         \centering
         \includegraphics[width=\textwidth]{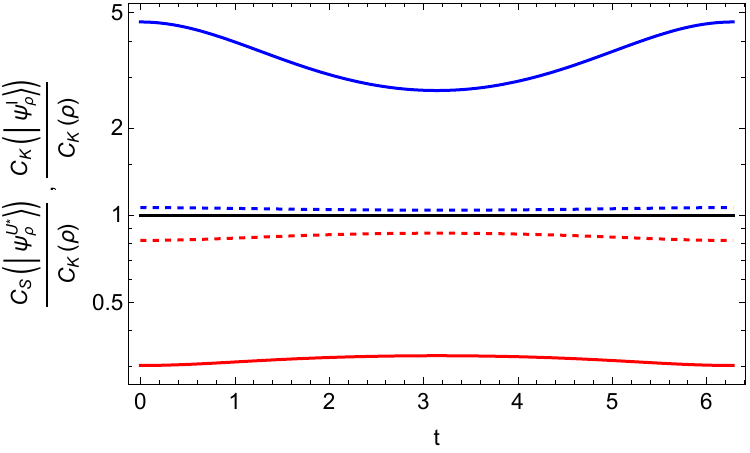}
         \caption{}
         \label{fig:skkk1q}
     \end{subfigure}
     \hfill
     \begin{subfigure}[b]{0.32\textwidth}
         \centering
         \includegraphics[width=\textwidth]{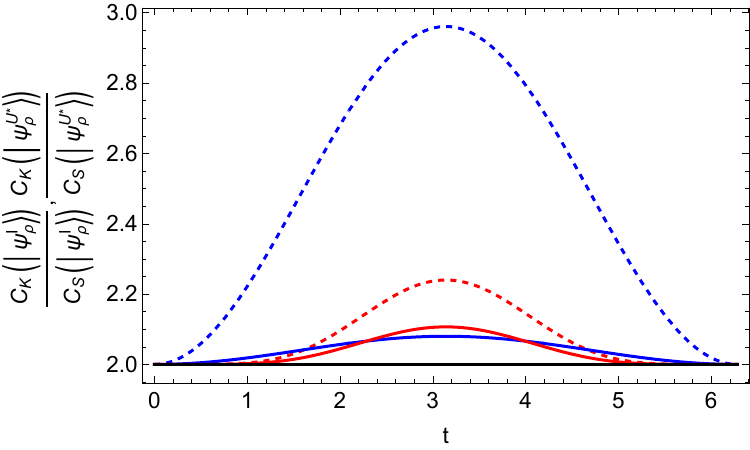}
         \caption{}
         \label{fig:ksks1q}
     \end{subfigure}
        \caption{Time dependence of the complexities of generic one-qubit mixed states~\eqref{eq:notation-1q} with different parameters $\Delta p=0.4,\theta=\pi/3$ (solid curve) and $\Delta p=0.98,\theta=\pi/4$ (dashed curve).
        (a)~Three complexities $\mathcal{C}_K^{\mathbb{I}}(t)$ (red), $\mathcal{C}_K(t)$ (blue), and $\mathcal{C}_S^{U^\ast}(t)$ (green).
        (b)~Ratio between the operator complexity and CoPs: 
        $\mathcal{C}_K^{\mathbb{I}}((t))/\mathcal{C}_K(t)$ (blue) and $\mathcal{C}_S^{U^\ast}(t)/\mathcal{C}_K(t)$ (red) in log plot. 
        (c)~Ratio $\mathcal{C}_K/\mathcal{C}_S$ between the state and operator CoPs for the time-independent purification
        (blue) and the time-dependent purification
        (red) with a reference line at $2$ (black).}
        \label{fig:complexity_1q}
\end{figure*} 

Following the moment method in the Krylov formalism, we obtain the following CoPs for generic one-qubit systems~\footnote{Note that the autocorrelation function should be normalized as $G(0)=1$ to compute the moments correctly.}:
\begin{align}
    C_K^{\,\mathbb{I}}(t) &= 2 \sin^2\frac{\tau}{2} \qty(1-(\Delta p)^2\cos^2\theta) \qty(1+\sin^2\frac{\tau}{2}(\Delta p)^2 \cos^2\theta)\\
    C_S^{\,\mathbb{I}}(t) &= \sin^2\frac{\tau}{2} \qty(1-(\Delta p)^2 \cos^2\theta) \\
    C_K(t) & = \frac{(\Delta p)^2 \sin^2\theta}{1+(\Delta p)^2}\qty(\sin^2 \tau + 8\qty(1-\frac{(\Delta p)^2\sin^2\theta}{1+(\Delta p)^2}) \sin^4\frac{\tau}{2})\\
    C_S^{U^\ast}(t) &= \frac{(\Delta \sqrt{p})^2 \sin^2\theta}{2} \qty(\sin^2 \tau + 8\qty(1-\frac{(\Delta \sqrt{p})^2\sin^2\theta}{2}) \sin^4 \frac{\tau}{2}) \\
    C_K^{U^\ast}(t) &= b_1^2 \qty[\qty(1-b_1^2\sin^2\frac{\tau}{2})^2\sin^2\tau + 8\sin^4\frac{\tau}{2}\qty[b_2^2-b_1^2 b_2^2 \sin^2\frac{\tau}{2} + b_1^2 \sin^4\frac{\tau}{2} \qty(\frac{(b_1^2-1)(2+(2-b_2^2)^2)}{b_2^2}+b_4^2(2-b_2^2))]]
\end{align}
where 
\begin{equation}
    b_1=\abs{\Delta\sqrt{p} \sin\theta}, \quad 
    b_2=\sqrt{1+\frac{1}{2}(\Delta \sqrt{p})^2 \sin^2\theta}, \quad 
    b_4=\sqrt{-14+3(\Delta\sqrt{p})^2\sin^2\theta + \frac{36}{2+(\Delta \sqrt{p})^2\sin^\theta}}
\end{equation}
are the Lanczos coefficients for $C_K^{U^\ast}(t)$.
It immediately follows that $C_K^{\mathbb{I}}\ge 2C_S^{\mathbb{I}}$ as $C_K^{\mathbb{I}}(t)/(2 C_S^{\mathbb{I}}(t))=1+\sin^2(\tau/2)(\Delta p)^2\cos^2\theta \ge 1$. We also confirmed $C_K^{U^\ast}\ge 2C_S^{U^\ast}$ analytically although it is less trivial. In Fig.~\ref{fig:complexity_1q}c, we show the ratio is indeed equal or above 2 over time for $\Delta p=0.4,\theta=\pi/3$ (solid lines) and $\Delta p=0.98,\theta=\pi/4$ as examples.

It is worth noting that $C_K^{\mathbb{I}},C_K,C_S^{U^\ast}$ all share the same Krylov structure, namely, there are only two non-vanishing Lanczos coefficients that correspond to the off-diagonal elements in the time evolution operator in the Krylov basis. This leads to the following form:
\begin{equation}
    C[\mu] = \mu \qty(\sin^2\tau + 8(1-\mu) \sin^4\frac{\tau}{2}).
\end{equation}
Corresponding to $C_K^{\mathbb{I}},C_K,C_S^{U^\ast}$,
\begin{equation}
    \mu = \mu^{\mathbb{I}}\equiv \frac{1-(\Delta p)^2 \cos^2\theta}{2}, \quad \mu_\ast\equiv \frac{(\Delta p)^2\sin^2\theta}{1+(\Delta p)^2}, \quad \mu^{U^\ast} \equiv \frac{(\Delta\sqrt{p})^2\sin^2\theta}{2}.
\end{equation}
From this expression, it is apparent that these complexities have a similar time dependence -- only differ in their coefficients.
Using $0\le p(1-p) \le 1/4$, one can show that the coefficients of $\sin^2\tau$ and $\sin^4(\tau/2)$ of $C_K^{\mathbb{I}}-C_K$ and $C_K-C_S^{U^\ast}$ are non-negative, proving our conjecture $C_K^{\mathbb{I}}\ge C_K\ge C_S^{U^\ast}$ for general unitarily evolving one-qubit mixed states as shown in Fig.~\ref{fig:complexity_1q}a. We also observe that the temporal fluctuations of the ratio $C_S^{U^\ast}(t)/C_K(t)$ is comparatively smaller compared to its mean value as shown in Fig.~\ref{fig:complexity_1q}b, confirming one of our conjectures.

\section{Mixed state and purified state complexity for different parameters}\label{sec:diff_ham}
\begin{figure*}[hbtp]
    \centering
     \begin{subfigure}[b]{0.45\textwidth} %previously 0.47
     \centering
         \includegraphics[width=\textwidth]{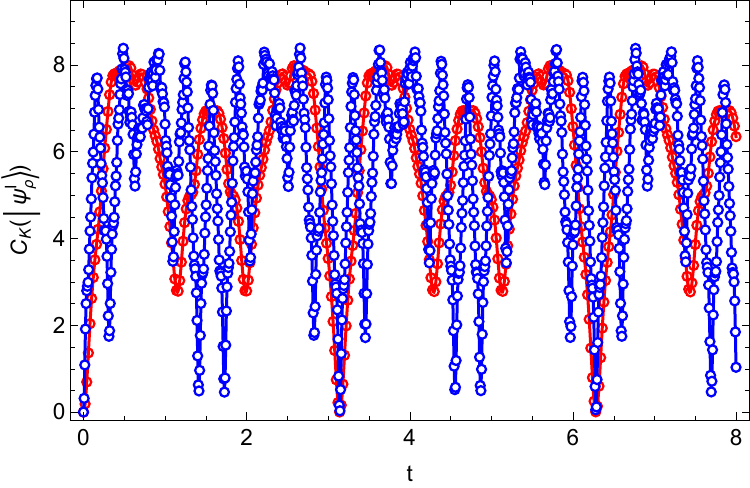}
         \caption{Krylov complexity of time-independent purification}
     \end{subfigure}
     \hfill
     \begin{subfigure}[b]{0.45\textwidth} %previously 0.47
         \centering
         \includegraphics[width=\textwidth]{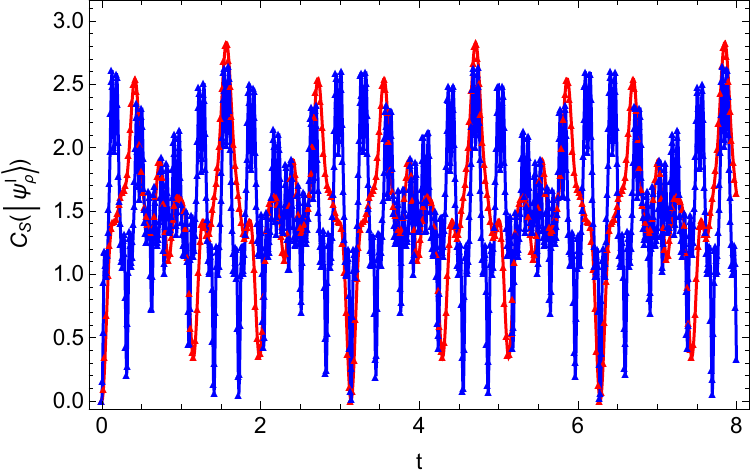}
         \caption{Spread complexity of time-independent purification}
     \end{subfigure}
        \caption{In both panels, the red color signifies the Hamiltonian parameters of~{(6)} %\eqref{eq:Hamiltonian_werner} 
        with $r=4$ and $q=15$, while the blue color signifies the Hamiltonian parameters with $r=10$ and $q=50$. (a) The empty circles, in both red and blue, signify the purified state with $p\rightarrow 1$, and the solid lines signify $p=1/4$ for the initial state~{(5)}. %\eqref{w_state}. 
        (b) The solid triangles, in both red and blue, signify the purified state with $p\rightarrow 1$, and the solid lines signify $p=1/4$. For both the operator and the state complexity, the complexity overlaps for the different purity parameters with the same Hamiltonian but differs for different Hamiltonian parameters. }
        \label{fig:TI_complexity}
    \centering
     \begin{subfigure}[b]{0.52\textwidth}
     \centering
         \includegraphics[width=\textwidth]{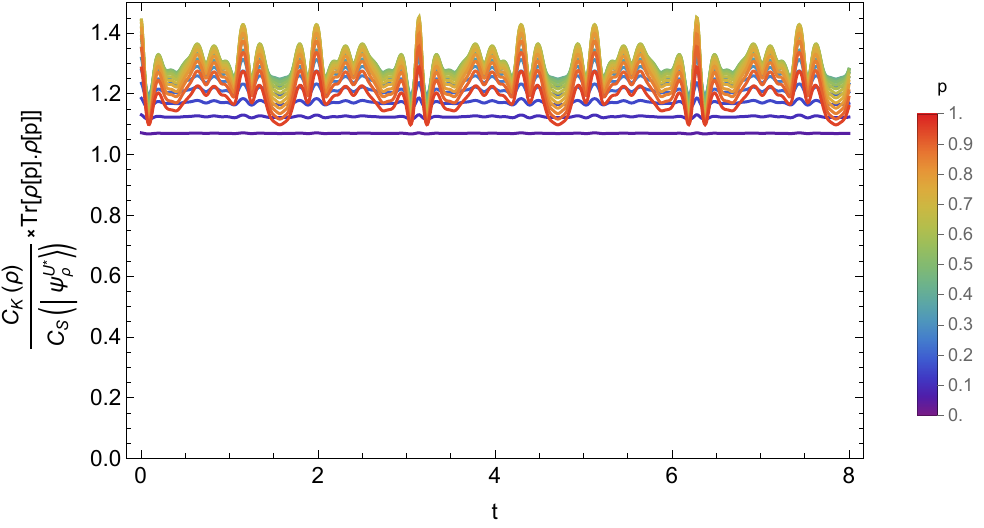}
         \caption{}
         \label{fig:ratio_purity}
     \end{subfigure}
     \hfill
     \begin{subfigure}[b]{0.45\textwidth}
         \centering
         \includegraphics[width=\textwidth]{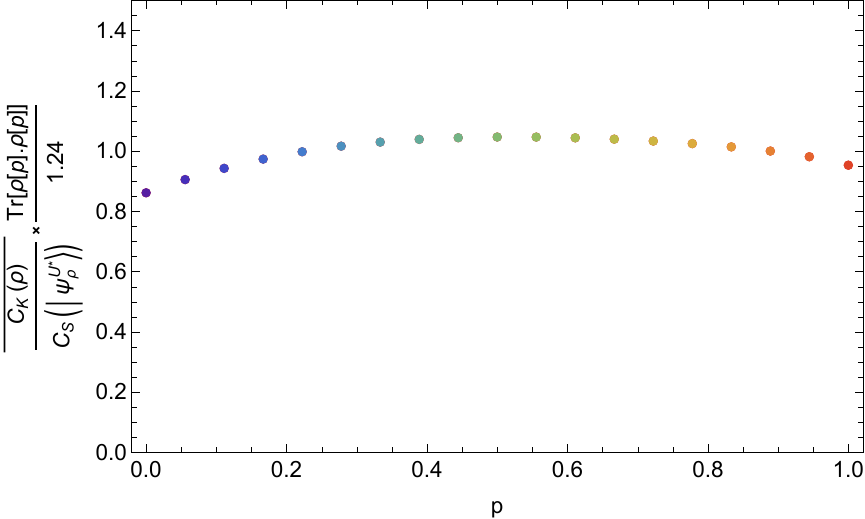}
         \caption{}
         \label{fig:avg_ratio_purity}
     \end{subfigure}
        \caption{The central legend bar showing different values of the parameter $p$ is common to both of the panels. The Hamiltonian~{(6)} %\eqref{eq:Hamiltonian_werner} 
        with $r=4$ and $q=15$ has been used to generate these results.  \hspace{\textwidth}
        (a) Time evolution of $R(\rho(t))$ for different values of purity of the initial state for the density matrix given by~{(5)}. %\eqref{w_state}
        $R(\rho(t))$ is concentrated around a particular range which shows that $\mathcal{C}_K(\rho(t))/\mathcal{C}_S\qty(\ket{\Psi_\rho^{U^\ast}(t)})$ is proportional to the inverse of the purity of the initial density matrix up to fluctuations.  \hspace{\textwidth}
        (b) Variation of the time average of the ratio times purity, $\overline{\mathcal{C}_K(\rho(t))/\mathcal{C}_S\qty(\ket{\Psi_\rho^{U^\ast}(t)})}\times\Tr(\rho^2)$.  
        }
        \label{fig:rat_pur}
    \hspace{\textwidth}\\ \hspace{\textwidth}
    \includegraphics[width=0.98\linewidth]{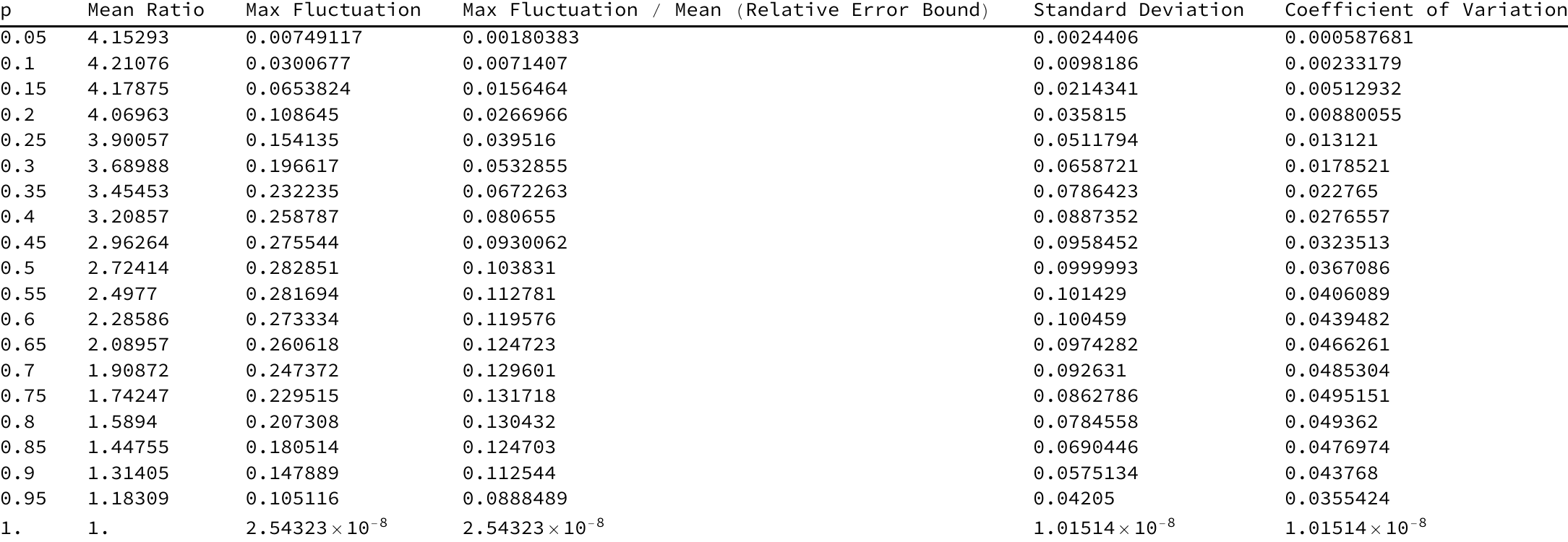}
    \caption{The fluctuation in the ratio of the complexities, $\mathcal{C}_K(\rho)/\mathcal{C}_S\qty(\ket{\Psi_\rho^{U^\ast}})$. The column "Mean Ratio" is the time average of $\mathcal{C}_K(\rho)/\mathcal{C}_S\qty(\ket{\Psi_\rho^{U^\ast}})$ over the period of recurrence time scale.}
    \label{fig:coefficient_of_variation_for_the_ratio}
\end{figure*}

 We observe that the complexities of time-independent purification, $\mathcal{C}_{S,K}(\ket{\psi_\rho^{\mathbb{I}}(t)})$, are almost insensitive to the purity of the initial state for Werner states. As we can see in Fig.~\ref{fig:TI_complexity}, in practice, $\mathcal{C}_{S,K}(\ket{\psi_\rho^{\mathbb{I}}(t)})$ overlaps for different values of purity of the initial state. This makes them a better probe for understanding the dynamical properties of the Hamiltonian. 
 Here, we show that the state/operator complexity of time-independent purification does indeed change for different parameters of the Hamiltonian but not for the different initial states. We propose for future studies that the complexity of time-independent purification could be used as a diagnostic for the time evolution, e.g. chaotic Hamiltonians exhibiting the transition from integrable to chaotic behaviour.
 
\section{Relation between $\mathcal{C}_K(\rho)/\mathcal{C}_S(|\Psi_\rho^{U^\ast}\rangle)$  and the purity of the initial mixed state}\label{sec:rat_pur}
Following~(4), %\eqref{eq:conjecture1}, 
the value of $\mathcal{C}_K(\rho)$ is lower bounded by $\mathcal{C}_S\qty(\ket{\Psi_\rho^{U^\ast}})$ and they behave quite similarly. In this part, we discuss the initial state dependence of the ratio $\mathcal{C}_K(\rho)/\mathcal{C}_S\qty(\ket{\Psi_\rho^{U^\ast}})$ is mostly given by the purity $\Tr\rho^2$~\footnote{Note that purity does not depend on time for a unitarily evolving density matrix.}. (Notice that the ratio considered in this section is the inverse of the ratio plotted in Fig.~2c%\ref{fig:ksks}
.) 
In Fig.~\ref{fig:ratio_purity}, we plot
\begin{equation}
    R(\rho(t)) := \frac{\mathcal{C}_K(\rho(t))}{\mathcal{C}_S\qty(\ket{\Psi_\rho^{U^\ast}(t)})}\times\Tr(\rho^2)
\end{equation}
as a function of time. We find that the ratio of the complexities, $R(\rho)$, is
inversely proportional to the purity of the initial state up to small fluctuations. This can be also seen from its temporal average over the recurrence time~\cite{Balasubramanian:2024ghv, Hashimoto:2023swv}, when complexity returns to zero, as shown in Fig.~\ref{fig:avg_ratio_purity}.

 Fig.~\ref{fig:coefficient_of_variation_for_the_ratio} lists various measures for the fluctuations of the ratio of complexities $\mathcal{C}_K(\rho)/\mathcal{C}_S(|\Psi_\rho^{U^\ast}\rangle)$ with different values of the parameter~$p$ in the Werner state~(5). %\eqref{w_state}. 
 We find that numerically, the state complexity of the time-dependent purification and the operator complexity of the original mixed states agree with each other reasonably well. For a quantitative evaluation, let us look at the coefficient of variation. In the current case, it is defined as the standard deviation over time divided by the average ratio over recurrence time. Fig.~\ref{fig:coefficient_of_variation_for_the_ratio} shows that the coefficient of variation is below 5\%, indicating the concentration around a constant value.

\bibliography{ref}
\end{document}